\documentclass[twocolumn]{jpsj2} 
\title{Anisotropic Behavior of Knight Shift in Superconducting State of Na$_x$CoO$_2$$\cdot~y$H$_2$O}

\author{
Yoshihiko \textsc{Ihara,}\thanks{E-mail address: ihara@scphys.kyoto-u.ac.jp},
Kenji \textsc{Ishida,}\thanks{E-mail address: kishida@scphys.kyoto-u.ac.jp},
Hideo \textsc{Takeya,}
Chishiro \textsc{Michioka,}$^{1}$
Masaki \textsc{Kato,}$^{1}$
Yutaka \textsc{Itoh,}$^{1}$
Kazuyoshi \textsc{Yoshimura,}$^{1}$
Kazunori \textsc{Takada,}$^{2}$
Takayoshi \textsc{Sasaki,}$^{2}$
Hiroya \textsc{Sakurai,}$^{3}$
and
Eiji \textsc{Takayama-Muromachi}$^{3}$}

\inst{
Department of Physics, Graduate School of Science, Kyoto University, Kyoto 606-8502, Japan \\
$^{1}$Department of Chemistry, Graduate School of Science, Kyoto University, Kyoto 606-8502, Japan \\
$^{2}$Advanced Materials Laboratory, National Institute for Materials Science, 1-1 Namiki, Tsukuba, Ibaraki 305-0044, Japan\\
$^{3}$Superconducting Materials Center, National Institute for Materials Science, 1-1 Namiki, Tsukuba, Ibaraki 305-0044, Japan
}

\abst{The Co Knight shift was measured in an aligned powder sample of Na$_x$CoO$_2$$\cdot~y$H$_2$O, which shows superconductivity at $T_c \sim 4.6$ K. The Knight-shift components parallel ($K_c$) and perpendicular to the $c$-axis (along the $ab$ plane $K_{ab}$) were measured in both the normal and superconducting (SC) states. The temperature dependences of $K_{ab}$ and $K_c$ are scaled with the bulk susceptibility, which shows that the microscopic susceptibility deduced from the Knight shift is related to Co-3$d$ spins. In the SC state, the Knight shift shows an anisotropic temperature dependence: $K_{ab}$ decreases below 5 K, whereas $K_c$ does not decrease within experimental accuracy. This result raises the possibility that spin-triplet superconductivity with the spin component of the pairs directed along the $c$-axis is realized in Na$_x$CoO$_2$$\cdot~y$H$_2$O.  
}

\kword{superconductivity, hydrate sodium cobalt oxide, NMR, spin fluctuations}

\begin{document}
\maketitle

Superconductivity in Na$_x$CoO$_2$$\cdot$$y$H$_2$O ($x\sim$ 0.35, $y\sim$ 1.3) with a transition temperature of $T_c \sim$ 5 K was discovered by Takada {\it et al} \cite{Takada}.
Although $T_c$ is smaller by one order of magnitude than in cuprate superconductors, much attention has been paid because of the unique two-dimensional layer crystal structure in which superconductivity occurs. 
The CoO$_2$ units form a two-dimensional hexagonal layered structure, which is in contrast with the tetragonal structure of cuprates.

Until now, various experiments revealed that this superconductivity is of a non-$s$-wave type, where electron correlations play an important role \cite{Ishida1,Fujimoto,Kanigel,Yang}. We have studied the relationship between magnetic fluctuations and the superconducting (SC) transition temperature $T_c$ \cite{Ihara1,Ihara2}, and shown that the highest $T_c$ in the system is observed in the vicinity of the magnetic phase \cite{Ihara2}. 
In addition, we suggest from the experimental point of view that the crystal-field splitting between $a_{1g}$ and $e'_g$ states of Co-$3d$ $t_{2g}$ orbitals is an important parameter for determining superconductivity from measurement of the electric quadrupole frequency $\nu_{\rm Q}$ by nuclear quadrupole resonance (NQR) on various samples with different values of $T_c$ \cite{Ihara2}.
Quite recently, a mechanism for the superconductivity in cobaltate superconductors was proposed by Yanase {\it et al.} \cite{Yanase} and Mochizuki {\it et al.} \cite{Mochizuki}, which is in good agreement with the above experimental results. In this mechanism, $p$-wave or $f$-wave spin-triplet superconductivity induced by ferromagnetic fluctuations is anticipated.
In fact, the possibility of spin-triplet superconductivity is suggested by several theoretical studies\cite{Ikeda,Kuroki,Khaliulin}.
In order to understand the mechanism of this superconductivity, experimental identification of the SC symmetry in Na$_x$CoO$_2$$\cdot$$y$H$_2$O is a crucial issue.

To identify the symmetry, we need to measure the spin susceptibility in the SC state. The most reliable and precise measurement of the spin susceptibility in the SC state is through the Knight shift, which measures the effective field at the nuclear site produced by conduction electrons. Although preliminary Knight-shift measurements using a powder sample show a decrease in the spin susceptibility for an applied field in the CoO$_2$ plane below $T_c$ \cite{Kobayashi1,Ihara3}, the symmetry of the SC pairing cannot be determined until the Knight shift for the field along the $c$-axis ($K_c$) is precisely measured. In this paper, we report results of Knight-shift measurements in the field parallel and perpendicular to the $c$-axis. It was found that $K_{c}$ does not decrease within an experimental accuracy in the SC state, which is in sharp contrast to that perpendicular to the $c$-axis ($K_{ab}$) in the SC state.

The powder sample we measured is carefully characterized by X-ray diffraction,  inductively coupled plasma atomic-emission spectroscopy (ICP-AES), and redox titration measurements. The results of the analyses are the following: the lattice parameters are $a$= 2.825 and $c$= 19.714 \AA, Na content is $x$ = 0.351 and the valence of Co is + 3.37. Oxygen atoms of the CoO$_6$ octahedron are partially exchanged by $^{17}$O atoms with the nuclear spin. The detailed $^{17}$O-NMR and $^{59}$Co-NQR results are already published in the literature. \cite{Ihara3} A magnetization measurement in a field of 1 mT shows the appearance of a Meissner signal below 4.6 K as shown in Fig. 1. The resonance frequency of the $\pm 5/2 \leftrightarrow \pm7/2$ transition of Co NQR is 12.45 MHz, slightly larger than that of the highest-$T_c$ sample (12.40 MHz).\cite{Ihara2} According to the phase diagram we developed,\cite{Ihara2} this sample is situated near the border between superconductivity and magnetism.
The nuclear spin-lattice lattice relaxation rate $1/T_1$ was measured by Co NQR. The temperature dependence of $1/T_1$ divided by temperature $1/T_1T$ shows an anomaly at 4.9 K, slightly higher than the onset temperature of the Meissner signal. The anomaly of $1/T_1T$ might be a SC precursor effect related to the two-dimensional character of the compound.
 
\begin{figure}[htbp]
\begin{center}
\includegraphics[width=8cm,clip]{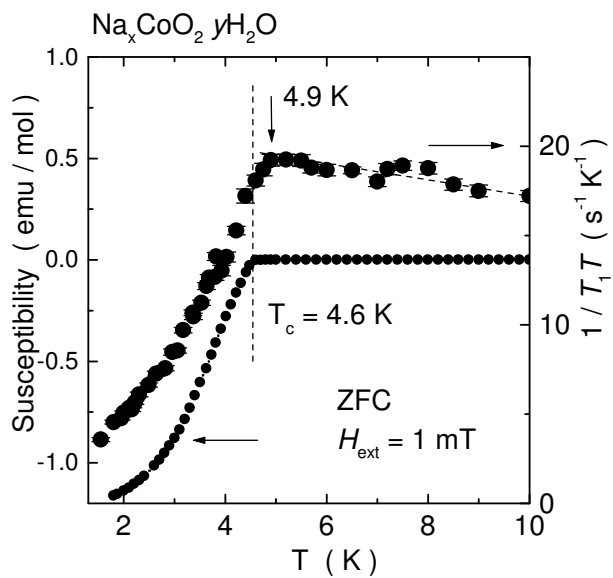}
\caption{Temperature dependence of $\chi_{\rm bulk}$ in Na$_x$CoO$_2$$\cdot~y$H$_2$O measured at 1 mT by zero-field cooling process. The temperature dependence of $1/T_1T$ is also shown. }
\label{spectra}
\end{center}
\end{figure}

The powder sample was aligned using the nonreactive liquid ``Fluorinert''in a magnetic field of 6 T and was fixed by solidifying the liquid as reported in the literature.\cite{Kato} In this condition, the crystalline $a$-and $b$-axes are parallel to the applied field due to the anisotropy of the magnetic susceptibility (i.e., the crystalline $c$-axis is distributed in the plane perpendicular to the field).\cite{Chou} The Co NMR spectrum obtained for this condition is denoted as a ``$H \| ab$ plane'' Co NMR spectrum, which is a two-dimensional powder pattern due to the distribution of the $a$- and $b$- axes with respective to the field.  
Then the aligned sample is rotated by 90$^{\circ}$ so that the plane in which the $c$-axis is distributed is parallel to the applied field. The Co-NMR spectrum taken under this condition is denoted as a ``$c$-axis powder-pattern'' spectrum. It should be noted that NMR signals arising from the $H \| c$ axis show peaks in the ``$c$-axis powder-pattern'' spectrum more clearly than those in an ordinary ``three-dimensional powder-pattern'' spectrum. The relationship between the anisotropy of the magnetization and NMR spectrum is precisely discussed in the literature. \cite{Young}

\begin{figure}[htbp]
\begin{center}
\includegraphics[width=8cm,clip]{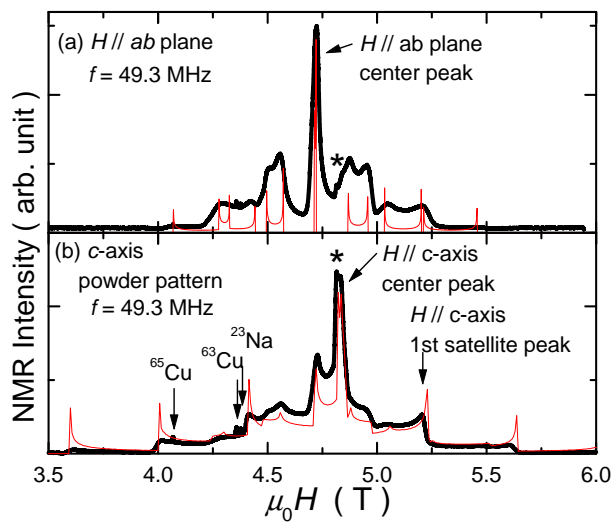}
\caption{Co NMR spectra for (a) $H \| ab$ plane and (b) $c$-axis powder pattern  in an aligned powder sample. The sharp peak shown by * arises from an impurity phase. Both powder-pattern spectra are well explained by calculation, shown by red lines. Here, the NMR parameters are $\nu_z=$4.10 MHz and $\eta=0.22$.    }
\label{spectra}
\end{center}
\end{figure}
Figure 2 shows the Co NMR spectra for the (a) $H \| ab$ plane and (b) $c$-axis powder-pattern, both of which were obtained at a frequency of 49.3 MHz by sweeping the magnetic field. Both spectra are well reproduced by calculations with NMR parameters in the figure caption. Spectrum (a) shows a typical two-dimensional powder pattern with a sharp peak (A) arising from the central ($1/2 \leftrightarrow -1/2$) transition. From peak (A), the Knight shift $K_{ab}$ is precisely determined. In spectrum (b), singularity peaks which are not observed in the $H \| ab$ plane spectrum are observed. These peaks arise from the $H \| c$-axis signal. However, a sharp peak marked by * which shows that the temperature-independent behavior overlaps with the $H \| c$ central-transition peak. This extra peak arises from some impurity phase, since this peak was observed even in the spectrum (a) and $T_1$ of this peak is extraordinarily long. Thus, we measured the Knight shift along the $c$-axis ($K_c$) at the 1st satellite peak (B) ($1/2 \leftrightarrow 3/2$ transition) of the $H \| c$-axis signal shown by the arrow, not at the central peak, although the signal intensity of the central peak is much stronger than that of other satellite peaks. 

\begin{figure}[htbp]
\begin{center}
\includegraphics[width=8cm,clip]{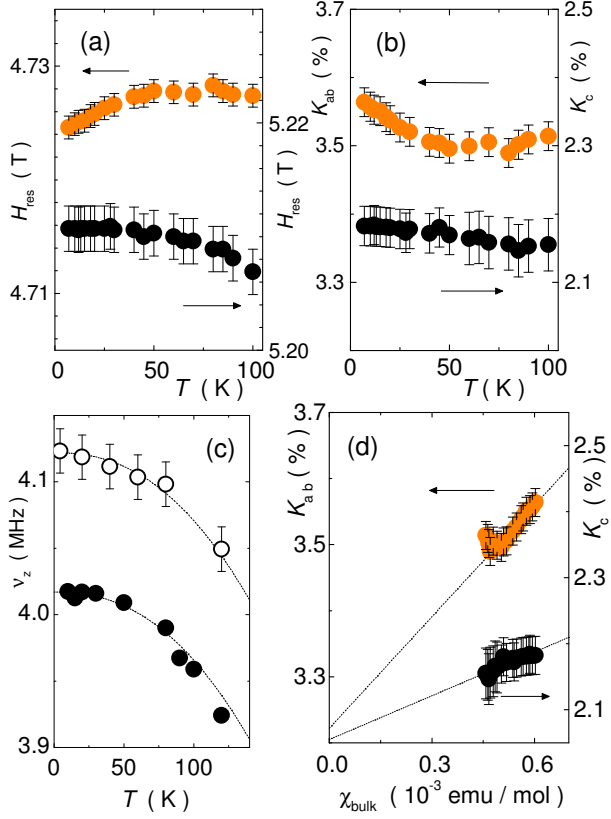}
\caption{(a) Temperature dependences of resonance fields of peaks (A) and (B) ($H_{\rm res}^{\rm A}$ and $H_{\rm res}^{\rm B}$). (b) Temperature dependences of $K_{ab}$ and $K_{c}$ estimated from $H_{\rm res}^{\rm A}$ and $H_{\rm res}^{\rm B}$ (see text). (c) Temperature dependences of $\nu_z$ values. Closed and open circles denote $\nu_z$ evaluated from two first-satellite peaks and two NQR transitions, respectively (see text). (d) $K_{ab}$ and $K_c$ are plotted against $\chi_{\rm bulk}$.}
\label{spectra}
\end{center}
\end{figure}
The resonance fields of peaks (A) and (B) ($H_{\rm res}^{\rm A}$ and $H_{\rm res}^{\rm B}$) are plotted in Fig. 3(a) against temperature. $K_{ab}$ and $K_c$ shown in Fig. 3(b) were evaluated from $H_{\rm res}^{\rm A}$ and $H_{\rm res}^{\rm B}$ using the following formula,
\begin{eqnarray*}
\gamma_{\rm n} H_{\rm res}^{\rm A} & = &\nu_0(1+K_{c})+\frac{15}{16}\frac{\nu_z^2}{\nu_0}\left(1-\frac{2}{3}\eta\right) \\
\gamma_{\rm n} H_{\rm res}^{\rm B} & = &\nu_0(1+K_{ab})+\nu_z-\frac{39}{72}\frac{\eta^2 \nu_z^2}{\nu_0},
\end{eqnarray*}
where $\gamma_{\rm n}$ and $\nu_0$ are the gyromagnetic ratio for a Co nucleus and the measured frequency, and $\nu_z$ and $\eta$ are the NQR frequency along the principle axis ($z$-axis) and the asymmetric parameter defined as $(\nu_{xx}-\nu_{yy})/\nu_{zz}$, respectively. 
In the estimation of the Knight-shift values using the NMR spectra with relatively larger NQR frequencies, the temperature dependence of the $\nu_{z}$ should be taken into account, and was measured by two independent methods, i.e., from a split of two satellite peaks in spectrum (b) of Fig. 2 and two NQR frequencies corresponding to the $\pm3/2 \leftrightarrow \pm5/2$ and $\pm5/2 \leftrightarrow \pm7/2$ transitions at various temperatures.
The $\nu_z$ values evaluated by these two methods are shown in Fig. 3(c).
The $\nu_z$ values estimated from spectrum (b) in Fig. 2 is slightly smaller than those from the NQR frequencies due to the ``pulling toward the center of the resonance'' which is typically seen in a powder pattern spectrum.
The temperature dependences are, however, in good agreement with each other. Using the $\nu_z$ values from a split of the satellite peaks, $K_c$ and $K_{ab}$ were estimated. $K_{ab}$ and $K_c$ are plotted against the bulk susceptibility in the temperature range between 10 and 100 K as shown in Fig. 3(d). The temperature dependences of $K_{ab}$ and $K_c$ are qualitatively different from the previous results showing a temperature-independent behavior,\cite{Ning1,Mukhamedshin,Ning2} but are in good agreement with the result by Kato {\it et al}. \cite{Kato} The Knight shift behavior seems to depend on samples, and is now being investigated in various samples.   

In metallic compounds, the observed $K_i(T)$ is the sum of two contributions: $K_i(T) = K_{\rm orb}^i +K_{s}^i(T)$ ($i = ab, c$), where $K_{\rm orb}^i$ is temperature-independent, and $K_s^i(T) = A_{\rm hf}^i\chi_s(T)$. Here, $A_{\rm hf}^i$ is a hyperfine coupling constant and $\chi_s(T)$ is the spin susceptibility. From the slope of Fig. 3(d), it is shown that the microscopic spin susceptibility at the Co nucleus site scales with the bulk susceptibility originating from Co-$3d$ spins through the coupling constants $A_{\rm hf}^{ab}$ and $A_{\rm hf}^c$, which are estimated to be $3.37 \pm 0.19$ and $1.21 \pm 0.11$ T /$\mu_B$, respectively.
The value of $A_{\rm hf}^{ab}$ is in good agreement with the previous value.\cite{Kato}
The anisotropy of $K_s^i$ is estimated to be $\varDelta K_c / \varDelta K_{ab} \sim 0.37 \pm 0.06$. The positive values of $A_{\rm hf}$ imply that for both $H \| c$ and $H \| ab$ the Fermi-contact interaction is dominant in the hyperfine coupling between the nucleus and electrons.
      
In the SC state, $\chi_s$ probes the spin symmetry of the Cooper pairs with a total spin of $S$ = 0 or 1. For spin-singlet pairing, $\chi_s$ in all directions decreases in the SC state, whereas for spin-triplet pairing, $\chi_s$ remains constant for the field parallel to the spin direction of the pair, but diminishes for the field perpendicular to the spin direction if the spin-orbit coupling is strong enough to pin the spin direction to a certain crystal axis. However, $\chi_s$ remains constant in all directions when the spin direction is changed by the applied field due to a pinning interaction smaller than the applied field. An anisotropic temperature dependence of $\chi_s$ is expected in the spin-triplet pairing state if the pinning interaction is stronger than the applied field.

\begin{figure}[htbp]
\begin{center}
\includegraphics[width=8cm,clip]{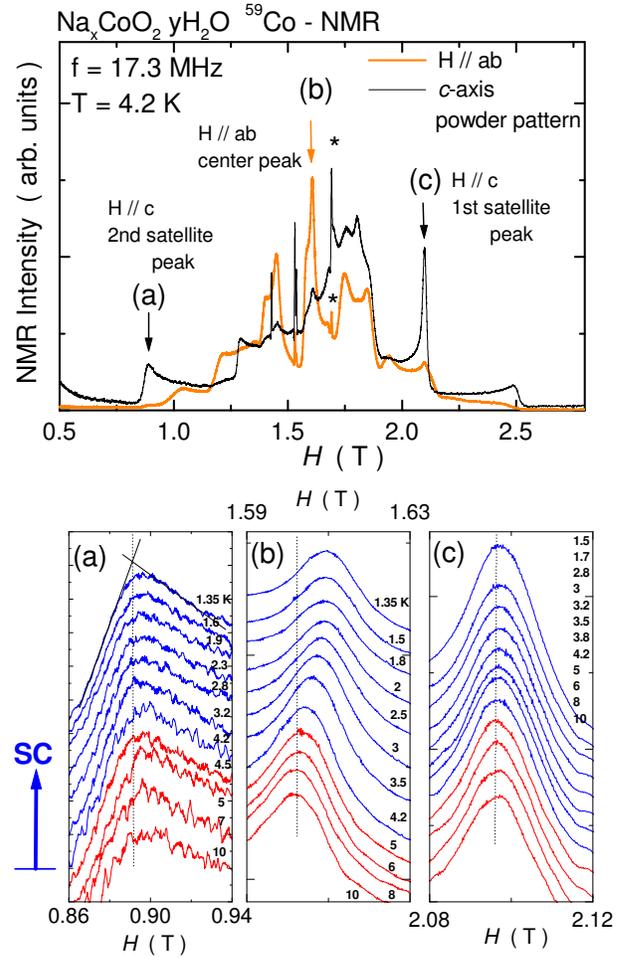}
\caption{Upper figure: Co NMR spectra of $H \| ab$ plane and $c$-axis powder pattern in aligned powder sample at a frequency of 17.3 MHz. The sharp peaks denoted by * arises from an impurity phase. Bottom figures: Temperature variation of peaks denoted by (a), (b) and (c). }
\label{spectra}
\end{center}
\end{figure}
The ``$H \| ab$ plane'' and ``$c$-axis powder-pattern'' Co NMR spectra obtained at a frequency of 17.3 MHz are shown in the upper panel of Fig. 4. 
The Knight shift below $T_c$ was measured at the several peaks shown by arrows in the spectra. 
Peak (b) in the spectra arising from the $H \| ab$ plane signal shifts below 5 K, but peaks (a) and (c) arising from the $H \| c$ signal do not shift, as shown in Fig. 4(b), and Fig. 4(a) and 4(c), respectively. 

\begin{figure}[htbp]
\begin{center}
\includegraphics[width=8cm,clip]{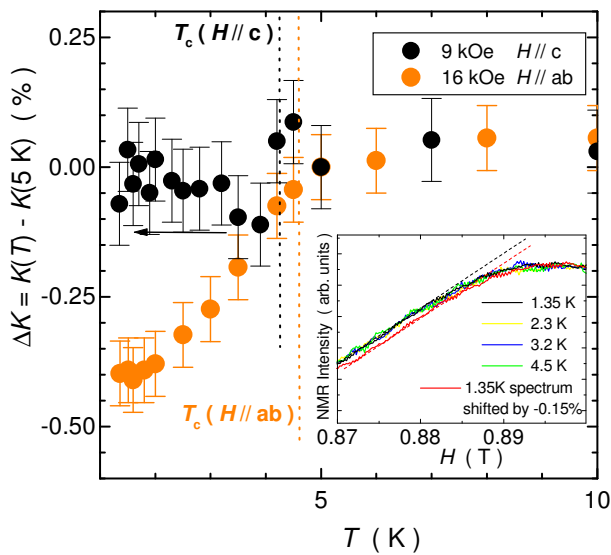}
\caption{Temperature dependences of $K_{ab}$ and $K_c$ determined from Fig. 4. In the figure, the deviation from the Knight shift at 5 K is plotted. The inset shows peak (a) measured at various temperatures. The 1.35 K spectrum shifted by $-0.15\%$ is also shown for reference (see text). }
\label{spectra}
\end{center}
\end{figure}
The temperature dependences of $K_{c}$ and $K_{ab}$ derived from the temperature variations of peaks (a) and (b) in Fig. 4 are shown in Fig. 5. Here, $K_{c}$ is determined at the cross point of two fitting lines between 0.865 and 0.885 T and between 0.900 and 0.920 T as shown in Fig. 4(a), and $K_{ab}$ is determined from the field at the peak maximum. 
In Fig. 5, the deviation from the Knight shift at $T = 5$ K is plotted against temperature. $K_{ab}$ decreases below 5 K slightly higher than $T_c$, which is consistent with the anomaly of $1/T_1T$ shown in Fig. 1. 
The $\varDelta K_{ab}$ at 1.5 K (= $K_{ab}$(1.5 K) $- K_{ab}$(5 K)) is $-0.4$\%.
If the SC pairing were in a spin-singlet state, $K_c$ should decrease down to $\varDelta K_c = -0.15 \pm 0.02$\%, which is estimated from the anisotropy of the normal-state $K_s^i$. This is because the spin component of the singlet pair diminishes in all directions. It should be noted that this change is one order of magnitude larger than the SC diamagnetic shift ($\sim 0.02$\%) which was observed from a sharp Na-NMR signal.\cite{IharaNa} Although the SC transition at 0.9 T parallel to the $c$-axis was confirmed by the measurement of $1/T_1$, such a decrease in $K_c$ is not observed at all. Peak (a) measured at various temperatures below 4.5 K is shown in the inset of Fig. 5, where the signal intensity of each spectrum is normalized. No shift is detectable in these spectra. The spectrum shifted by $-0.15$\%, which is the expected Knight-shift decrease in the case of the singlet pairing, is also shown in the inset by the red line. Such a small difference is difficult to detect from the peak position, but can be recognized from the steep slope on the left as seen in the inset. We stress that the change of $-0.15$\% is detectable from the present experimental accuracy, but was not observed in the measurement. Our result is in sharp contrast to the previous $K_c$ result reported by Kobayashi {\it et al.}\cite{Kobayashi2}, in which single crystals of Na$_x$CoO$_2$$\cdot~1.3$H$_2$O including non-SC compounds were used. The reasons for the discrepancy are unclear at the moment, but should be clarified in the near future. Further Knight-shift measurements using a high-quality single crystal and/or a c-axis-aligned sample are highly desired.  

The behavior of $\chi_s$ in the SC state revealed by the present Knight-shift measurement cannot be explained by a spin-singlet pairing state, but raises the possibility of a spin-triplet superconductivity with the spin component of the SC pairs parallel to the $c$-axis. One of the most promising states within such a spin-triplet state is expressed as $\hat{\mbox{\boldmath$x$}}k_x+\hat{\mbox{\boldmath$y$}}k_y$ using the SC {\bf d}-vector. This state is analogous to the $B$-phase of superfluid $^3$He, which does not break time-reversal symmetry. 
The absence of the spontaneous field revealed by the $\mu$SR measurements is consistent with this state.\cite{Higemoto} 

In conclusion, we have shown from Knight-shift measurements using an aligned powder sample that the temperature dependence of $\chi_s$ in the SC state of Na$_x$CoO$_2$$\cdot$$y$H$_2$O is anisotropic: $\chi_{s}^{ab}$ decreases below 5 K but $\chi_{s}^{c}$ does not within an experimental accuracy. This result raises the possibility that Na$_x$CoO$_2$$\cdot$$y$H$_2$O is a spin-triplet superconductor, in which the spin component of the triplet pair is parallel to the $c$-axis.

We thank D.~E. MacLaughlin, H. Yaguchi, S. Nakatsuji and Y.~Maeno for experimental support and valuable discussions. We also thank H.~Ikeda, S.~Fujimoto, K.~Yamada, Y.~Yanase, M. Mochizuki, and M.~Ogata for valuable discussions.
This work was partially supported by CREST of the Japan Science and Technology Agency (JST) and the 21 COE program on ``Center for Diversity and Universality in Physics'' from MEXT of Japan, and by Grants-in-Aid for Scientific Research from the Japan Society for the Promotion of Science (JSPS)(No.16340111) and MEXT(No.16076209).


\begin{thebibliography}{99}

\bibitem{Takada}
K.~Takada, H.~Sakurai, E.~Takayama-Muromachi, F.~Izumi, R.~A.~Dilanian and T.~Sasaki: Nature {\bf 422} (2003) 53.

\bibitem{Ishida1}
K.~Ishida, Y.~Ihara, Y.~Maeno, C.~Michioka, M.~Kato, K.~Yoshimura, K.~Takada, T.~Sasaki, H.~Sakurai and E.~Takayama-Muromachi: J. Phys. Soc. Jpn. {\bf 72} (2003) 3041.

\bibitem{Fujimoto}
T.~Fujimoto, G-q.~Zheng, Y.~Kitaoka, R.~L.~Meng, J.~Cmaidalka and C.~W.~Chu: Phys Rev Lett. {\bf 92} (2004) 047004.

\bibitem{Kanigel}
A.~Kanigel, A.~Keren, L.~Patlagan, K.~B.~Chashka and P.~King: Phys. Rev. Lett. {\bf 92} (2004) 257007.

\bibitem{Yang}
H.D.~Yang, J.-Y.~Lin, C.P.~Sun, Y.C.~Kang, C.L.~Huang, K.~Takada, T.~Sasaki, H.~Sakurai and E.~Takayama-Muromachi: Phys. Rev. B {\bf 71} (2005) 020505.

\bibitem{Ihara1}
Y.~Ihara, K.~Ishida, C.~Michioka, M.~Kato, K.~Yoshimura, K.~Takada, T.~Sasaki, H.~Sakurai, and E.~Takayama-Muromachi: J. Phys. Soc. Jpn. {\bf 73} (2004) 2069.

\bibitem{Ihara2}
Y.~Ihara, K.~Ishida, C.~Michioka, M.~Kato, K.~Yoshimura, K.~Takada, T.~Sasaki, H.~Sakurai, and E.~Takayama-Muromachi: J. Phys. Soc. Jpn. {\bf 74} (2005) 867.

\bibitem{Yanase}
Y.~Yanase, M.~Mochizuki and M.~Ogata: J. Phys. Soc. Jpn. {\bf 74} (2005) 430.

\bibitem{Mochizuki}
M.~Mochizuki, Y.~Yanase, and M. Ogata: Phys. Rev. Lett: {\bf 94} (2005) 147005.

\bibitem{Ikeda}
H.~Ikeda, and Y.~Nisikawa and K.~Yamada, J. Phys. Soc. Jpn. {\bf 73} (2004) 17.

\bibitem{Kuroki}
K.~Kuroki, Y.~Tanaka  and R.~Arita: Phys. Rev. Lett. {\bf 93} (2004) 077001 1-4.
\bibitem{Khaliulin}
G.~Khaliulin, W.~Koshibae, and S. Maekawa: Phys. Rev. Lett. {\bf 93} (2004) 176401.

\bibitem{Kobayashi1}
Y.~Kobayashi, M.~Yokoi, and M.~Sato: J. Phys. Soc. Jpn. {\bf 72} (2003) 2453. 

\bibitem{Ihara3}
Y.~Ihara, K.~Ishida, K.~Yoshimura, K.~Takada, T.~Sasaki, H.~Sakurai, and E.~Takayama-Muromachi: to be published in J. Phys. Soc. Jpn. {\bf 74} (2005).

\bibitem{Kato}
M. Kato {\it et al.}: cond/mat 0306036. 

\bibitem{Chou}
F.C.~Chou, J.H.~Cho, P.A.~Lee, E.T.~Abel, K.~Matan, and Y.S.~Lee: Phys. Rev. Lett. {\bf 92} (2004) 157004.

\bibitem{Young}
B.~-L. Young {\it et al.}: Rev. Sci. Ins. {\bf 73} (2002) 3038. 

\bibitem{Ning1}
F.~L.~Ning, T.~Imai, B.~W.~Statt, and F.~C.~Chou: Phys. Rev. Lett. {\bf 93} (2004) 237201.

\bibitem{Mukhamedshin}
I.R.Mukhamedshin, H.Alloul, G. Collin, and N. Blanchard: Phys. Rev. Lett. {\bf 94} (2005) 247602.

\bibitem{Ning2}
F.L.Ning and T. Imai: Phys. Rev. Lett. {\bf 94} (2005) 227004.

\bibitem{IharaNa}
Y.~Ihara: unpublished.

\bibitem{Kobayashi2}
Y. Kobayashi, H. Watanabe, M. Yokoi, T. Moyoshi, Y. Mori, and M. Sato: J. Phys. Soc. Jpn. {\bf 74} (2005) 1800.

\bibitem{Higemoto}
W.~Higemoto, K.~Ohishi, A.~Koda, S.~R.~Saha, R.~Kadono, K.~Ishida, K.~Takada, H.~Sakurai, E.~Takayama-Muromachi, and T.~Sasaki: Phys. Rev. B {\bf 70} (2005) 134508.







\end{thebibliography}
\end{document}